\documentclass[final,5p,times,twocolumn]{elsarticle}  

\usepackage{hyperref}

\journal{Journal of \LaTeX\ Templates}

\usepackage{amsmath}
\usepackage{amssymb}

\begin{document}

\begin{frontmatter}

\title{Advances Towards a Large-Area, Ultra-Low-Gas-Consumption RPC Detector}


\author[mymainaddress]{J. Saraiva\corref{mycorrespondingauthor}}

\cortext[mycorrespondingauthor]{Corresponding author}
\ead{joao.saraiva@coimbra.lip.pt}

\author[mysecondaryaddress]{C. Alemparte}
\author[mysecondaryaddress]{D. Belver}
\author[mymainaddress]{A. Blanco}
\author[mysecondaryaddress]{J. Call\'{o}n}
\author[mysecondaryaddress]{J. Collazo}
\author[mysecondaryaddress]{A. Iglesias}
\author[mymainaddress]{L. Lopes}

\address[mymainaddress]{LIP, Laborat\'{o}rio de Instrumenta\c{c}\~{a}o e F\'{i}sica Experimental de Part\'{i}culas, 3004-516 Coimbra, Portugal}

\address[mysecondaryaddress]{Hidronav Technologies SL, 36202 Vigo, Pontevedra, Spain}

\begin{abstract}
Large Resistive Plate Chamber systems have their roots in High Energy Physics experiments at the European Organization for Nuclear Research: ATLAS, CMS and ALICE, where hundreds of square meters of both trigger and timing RPCs have been deployed. These devices operate with complex gas systems, equipped with re-circulation and purification units, which require a fresh gas supply of the order of 6 cm$^{3}$/min/m$^{2}$, creating logistical, technical and financial problems.

In this communication, we present a new concept in the construction of RPCs which allowed us to operate a detector at ultra-low gas flow regime. With this new approach, the glass stack is encapsulated in a tight plastic box made of polypropylene, which presents excellent water vapor blocking properties as well as a good protection against atmospheric gases. 

\end{abstract}


\end{frontmatter}


\section{Introduction}

Large Resistive Plate Chamber (RPC) systems have their roots in High Energy Physics (HEP) experiments at the European Organization for Nuclear Research (CERN): ATLAS, CMS and ALICE, where hundreds of square meters of both trigger and timing RPCs have been deployed. These devices operate with complex gas systems, equipped with re-circulation and purification units, which require a fresh gas supply of the order of 6 cm$^{3}$/min/m$^{2}$, creating logistical, technical and financial problems~\cite{closedLoopGasSystem}. Recently, new EU legislation (nº 517/2014) for the progressive phasing out of the main gas used on RPCs - the 1,1,1,2-tetrafluoroethane (C$_{2}$H$_{2}$F$_{4}$) or R-134a - due to its Global Warming Potential over 100 years (GWP$_{100}$) of 1430, has further increased constraints on these systems.

In this communication, we present a new concept in the construction of RPCs which allowed us to operate a detector at ultra-low gas flow regime. With this approach, the glass stack is encapsulated in a tight plastic box made of polypropylene, which presents excellent water vapor blocking properties as well as a good protection against atmospheric gases. As shown in this paper, a detector with almost 2 m$^{2}$ was constructed and operated for more than one month with a gas flux of 1 cm$^{3}$/min/m$^{2}$ in stable conditions.

\section{Detector, Front-End \& DAQ}

The detector here presented uses a novel concept on the construction of RPCs already introduced in \cite{Blanco_2015, Lopes_2019}. With this approach, two sensitive parts of an RPC, the gas volume and the High Voltage (HV) insulation, are confined inside a permanent sealed plastic box. In order to operate the detector at ultra-low gas flux regime and, by this way, overcoming the previously mentioned phasing out of the R-134a, special care was paid to the material used for the gas system, including the sealed plastic box. Trying to improve the tightness against contaminants such as water vapor and inspired by the food packaging industry, we used, for the first time, polypropylene (PP) as enclosure of the gas volume, instead of a combination of acrylic (PMMA) and polycarbonate (PC) for the box and its lid, respectively. With this approach, the RPC could be operated in stable conditions during more than one month with roughly $1$~cm$^3$/min/m$^2$ of C$_2$H$_2$F$_4$ as described below.

With an active area of 1.2x1.6 m$^2$, the detector consists of two plastic boxes (two modules) each containing a multi-gap RPC with, between them, a printed circuit board of 64 longitudinal signal pick-up strips, 1.55 cm wide and 1.85 cm pitch.

\begin{figure}[ht]
\centering
\includegraphics*[width=80mm]{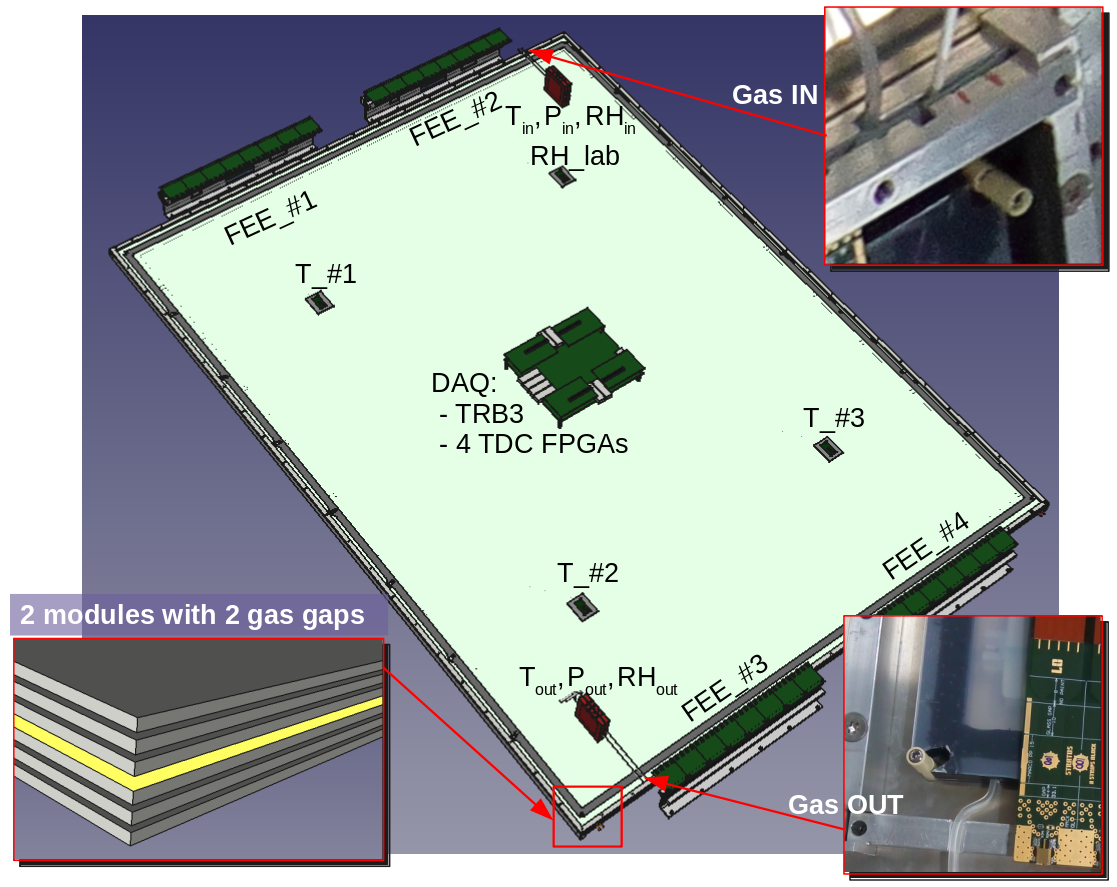}
\caption{Details of the detector; top-right: gas inlet; bottom-right: gas outlet; bottom-left: two sensitive volumes, with two gas gaps each, separated by the readout PCB (not shown for better understanding but with its location represented by the yellow color).}
\label{rpc}
\end{figure}

Additional details of the detector are provided in Figure~\ref{rpc}, namely:
\begin{itemize}
\item two multi-gap RPC, one on each side of the pick-up strips, with two gas gaps each, 1 mm wide, and glass 2 mm thick with a bulk resistivity of $\approx$ 4x10$^{12}$~$\Omega$cm at $25$~$^\circ$C;
\item the Data Acquisition System (DAQ), based on a multi-purpose Trigger Readout Board (TRB3)~\cite{trb3-ref}, using four FPGA-based TDCs with time resolution close to 20 ps $\sigma$;
\item the front-end electronics (FEE)~\cite{fee-ref} with time resolution around 35 ps $\sigma$;
\item environmental sensors: three TMP75 sensors for temperature (T) measurements and one SHT31 for the relative humidity (RH\_lab);
\item temperature, atmospheric pressure and humidity sensors of the gas at both, inlet and outlet sides of the RPC: two BME280 sensors, inserted in two little boxes, made of PP, through which the gas entering and exiting the RPC flows: T\textsubscript{in/out}, P\textsubscript{in/out} and RH\textsubscript{in/out}, respectively.
\end{itemize}

The gas monitoring sensors, located outside the metallic cage, allowed us to measure the R-134a flux exiting the detector. Moreover, in order to keep the gas gain constant along the whole period of test, the HV supplied to both modules was constantly modulated as a function of the environmental pressure and temperature, as described here~\cite{hv-ref}.

\section{Time, Charge and 2D Position}
Cosmic rays, such as muons of few GeV, crossing the RPC, create avalanches in the gas-filled regions which movement toward the electrodes induces current pulses in the strips. These pulses are fed to pre-amps located at both sides of the detector, as shown in Figure~\ref{rpc}.
The time and charge of these signals are afterwards encoded, respectively, in the leading edge and the width of the LVDS signals sent from the FEE to the TDCs.

Based on the time and charge information of the signals collected at both extremities of the strips ‒ designed here as front (F) and back (B) ‒, one can compute, for each event: its time (($T_F+T_B)/2$) and the respective charge (($Q_F+Q_B)/2$). Furthermore, the position of the event is obtained as follows: the strip with higher charge provides the position in one dimension, while ($T_F-T_B)/2$) gives the position in the other dimension (along the strip).

\section{Low Flux Operation}

\subsection{One month operation at low flux}

\begin{figure}[ht]
\centering
\includegraphics*[width=80mm]{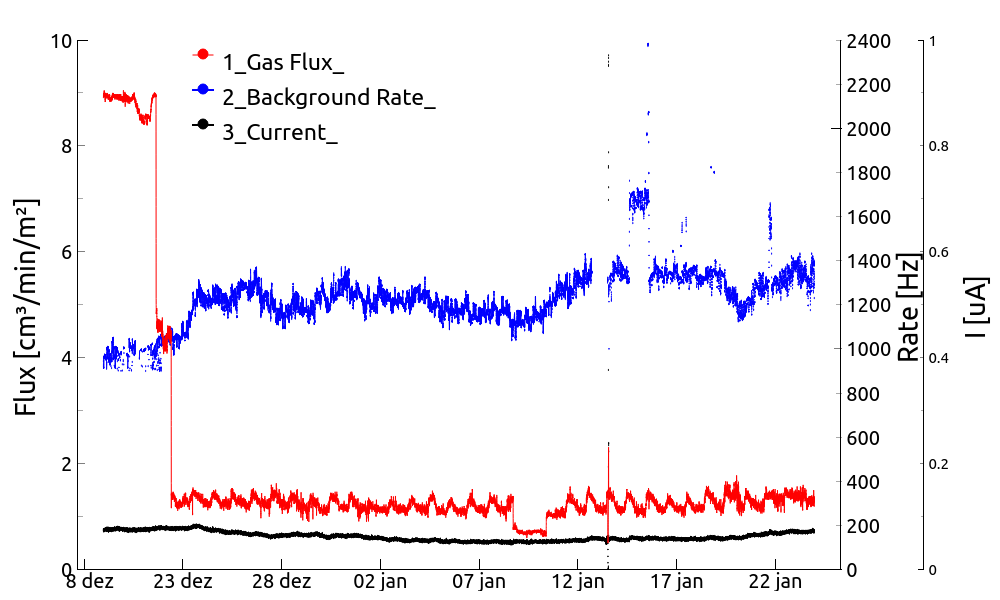}
\caption{One month-long operation of the RPC at very low gas flux (slightly above 1 cm$^{3}$/min/m$^{2}$) (left y-axis) and the respective current and background rate (right y-axis) showing that the RPC was operating under stable conditions.}
\label{oneMonthLowFlux}
\end{figure}

The RPC was operated in stable conditions from December 22 2021 to January 26 2022, with an R-134a flow around 2 cm$^{3}$/min, which corresponds, taking into account to the dimensions of its active area (1.2x1.6 m$^{2}$), to a gas flux around 1 cm$^{3}$/min/m$^{2}$ (Figure~\ref{oneMonthLowFlux}).

This test allowed us to conclude that the innovative solutions developed in this project, such as the choice of PP as enclosure material of the RPC sensitive volume and the rest of the gas system, resulted in a significant stabilization of the gas volume, guaranteeing high sealing performance against external contaminants, even at such low gas flow regime.

\subsection{Opeartion at residual flux}

At the end of the previously mentioned test at very low flux, a further reduction of the R-134a flux was performed on January 26 2022, reducing it from 1 cm$^{3}$/min/m$^{2}$ to almost zero.
As a result, an instantaneous increase of the background rate was observed, rising up to 50\% during the five days following the flux reduction. The values of relative humidity (RH) of the gas entering and exiting the RPC also presented an instantaneous increase just after the gas reduction: around 40\% and 10\% in the first day at the inlet and outlet sides of the RPC, respectively.

\subsection{Gas-flux re-establishment}

\begin{figure}[ht]
\centering
\includegraphics*[width=82mm]{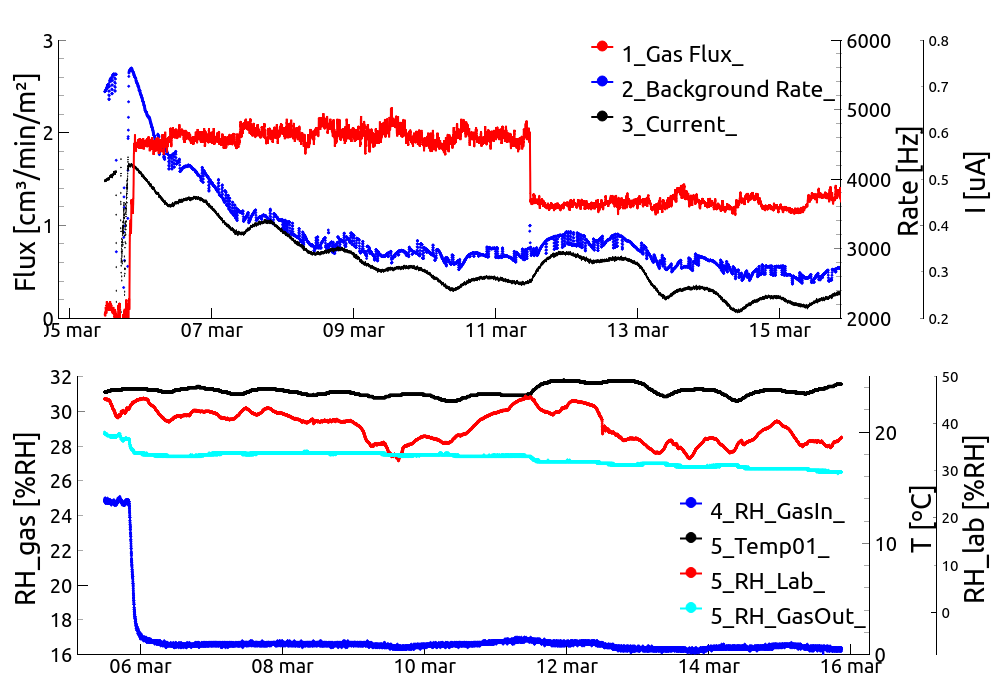}
\caption{At the top, while the gas flux is re-established, the background rate decreases by half in few days; at the bottom, instantaneous and simultaneous reduction of the RH of the gas entering and leaving the RPC when the flux is restored.}
\label{fifteenDaysLowFlux}
\end{figure}

One month later, the gas flux was brought back to values close to the initial situation: 1‒2 cm$^{3}$/min/m$^{2}$. As a result, the background rate and the provided current decreased significantly as shown in Figure~\ref{fifteenDaysLowFlux}. 

The results of the previous tests indicate that (1) the RPC can be operated at very low flux, and (2) even after a long period of time with residual flux, the RPC can gradually return to its previous state.

\section{Analysis at residual flux}

In order to identify the regions of the detector that contributed the most to the increase of the background rate during the days following the flux reduction, an additional analysis of the RPC parameters was performed using the following approach:

\begin{itemize}
\item split the area corresponding to the sensitive region of the RPC into 5x5 cm$^2$ bins (32x24 bins in total);
\item compute values per bin of the following quantities: background rate and deposited charge;
\item evaluate temporal and spatial evolutions of the mentioned quantities performing a bin-wise division between corresponding values spaced out over time.
\end{itemize}

\begin{figure}[ht]
\centering
\includegraphics*[trim={0.6cm 0.1cm 0.6cm 0.7cm},clip, width=68mm]{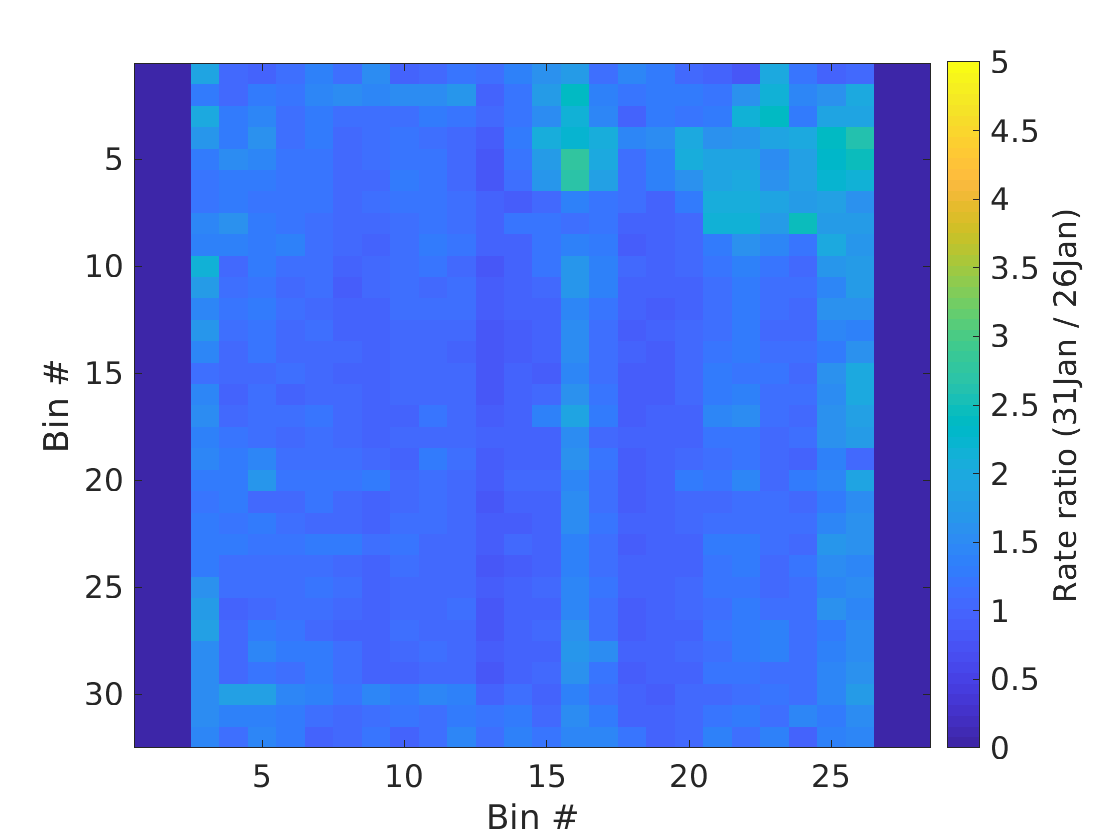}
\caption{Bin-wise division of the background rate of the RPC spaced out over 5 days after the gas flow cut off. A ratio above one means that the plotted quantity increased with time in this bin.}
\label{cosmicHitsPerBin_31JanVs26Jan}
\end{figure}

\begin{figure}[ht]
\centering
\includegraphics*[trim={0.6cm 0.1cm 0.6cm 0.7cm},clip, width=68mm]{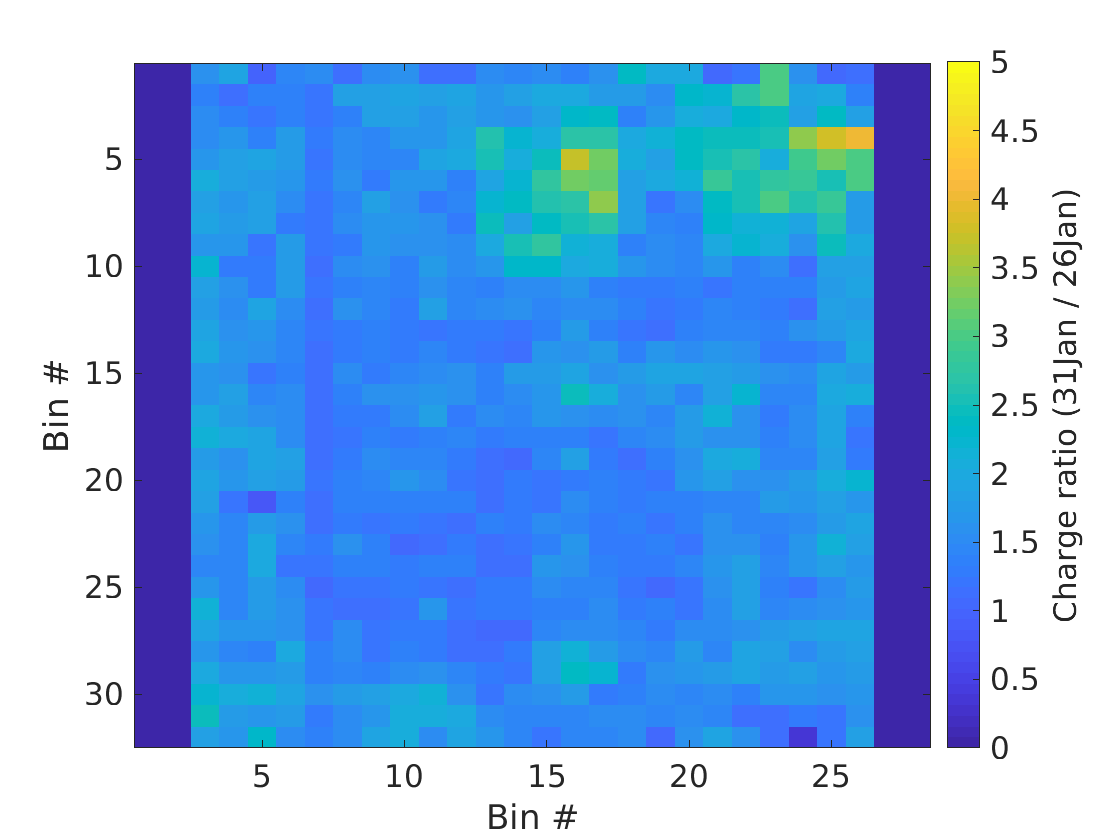}
\caption{Bin-wise division of the charge collected by the RPC spaced out over the same period of time than previous figure.}
\label{QPerBin_31JanVs26Jan}
\end{figure}

The 2D plots presented in this section show the temporal evolution obtained dividing the quantities per bin of January 31 by the same quantities obtained five days before, on January 26, corresponding to the day of the above-mentioned flow reduction. In this way, it was possible to visualize the evolution of the computed quantities and the most affected regions in the RPC. Figure~\ref{cosmicHitsPerBin_31JanVs26Jan} refers to the ratio of the background rate of the RPC during the mentioned period of time and Figure~\ref{QPerBin_31JanVs26Jan} shows the respective charge variation. From the analysis of these figures it seems clear that the most affected region of the RPC was the one close to the gas inlet. The same region also presented the higher increase of RH. In contrast, the region of the RPC close to the gas outlet might have been slightly affected but it is hard to confirm it based on the data collected and shown in these plots.

These results allowed us to conclude that the 50\% increase of the background rate occurred mainly in a circumscribed region close to the gas inlet of the RPC. Furthermore, the quick rise of both RH of the gas entering and exiting the RPC might be due to water vapor contamination near both sensors simultaneously, probably due to a lack of sealing in these measurement points.

\section{Conclusion}

In this communication, we presented a new concept in the construction of RPCs enabling operation at ultra-low gas flow regime. Among other improvements, the encapsulation of the sensitive volume within a tight plastic box made of polypropylene, which presents excellent water vapor blocking properties as well as a good blocking to atmospheric gases, allowed us to operate an RPC of almost 2 m$^2$ for more than one month with a gas flux around 1 cm$^{3}$/min/m$^2$. We also showed that decreasing even more the gas flux to a residual value, the sealing of the polypropylene box to contaminants was still effective, however, contamination arose from outside the RPC, upstream and downstream. One month later, re-establishing the gas flux, the RPC was able to recover its normal operation.

\section{Acknowledgements}
This work was supported by Fundação para a Ciência e Tecnologia, Portugal (project CERN/FIS-INS/0009/2019).


\end{document}